\documentclass[preprint,floatfix,superscriptaddress]{revtex4}
\usepackage{graphicx}
\usepackage{bbm}
\usepackage{amsmath,amssymb}
\newcommand{\be}{\begin{equation}}
\newcommand{\beq}{\begin{equation}}
\newcommand{\ee}{\end{equation}}
\newcommand{\bea}{\begin{eqnarray}}
\newcommand{\eea}{\end{eqnarray}}
\newcommand{\ba}{\begin{array}}
\newcommand{\ea}{\end{array}}

\renewcommand{\vr} {{\bf r}}
\newcommand{\vj} {{\bf j}}

\renewcommand{\vr} {{\bf r}}

\begin{document}
\title{Universal correction for the Becke-Johnson exchange potential}
\author{E. R{\"a}s{\"a}nen}
\email[Electronic address:\;]{erasanen@jyu.fi}
\affiliation{Nanoscience Center, Department of Physics, University of
  Jyv\"askyl\"a, FI-40014 Jyv\"askyl\"a, Finland}
\author{S. Pittalis}
\affiliation{Department of Physics and Astronomy, University of Missouri, Columbia, Missouri 65211, USA}
\affiliation{Institut f{\"u}r Theoretische Physik,
Freie Universit{\"a}t Berlin, Arnimallee 14, D-14195 Berlin, Germany}
\affiliation{European Theoretical Spectroscopy Facility (ETSF)}
\author{C. R. Proetto}
\altaffiliation[Permanent address: ]{Centro At{\'o}mico Bariloche and Instituto Balseiro, 8400
S.C. de Bariloche, R{\'i}o Negro, Argentina}
\affiliation{Institut f{\"u}r Theoretische Physik,
Freie Universit{\"a}t Berlin, Arnimallee 14, D-14195 Berlin, Germany}
\affiliation{European Theoretical Spectroscopy Facility (ETSF)}

\date{\today}

\begin{abstract}
The Becke-Johnson exchange potential [J. Chem. Phys. {\bf 124},
221101 (2006)] has been successfully used in electronic structure 
calculations within density-functional theory. However, in 
its original form the potential may dramatically fail in systems with 
non-Coulombic external potentials, or in the presence of
external magnetic or electric fields. 
Here, we provide a system-independent correction to the Becke-Johnson
approximation by (i) enforcing its
gauge-invariance and (ii) making it exact 
for any single-electron system. The resulting approximation is then
better designed to deal with current-carrying states, 
and recovers the correct asymptotic behavior for systems with any number of electrons. 
Tests of the resulting corrected exchange potential show very good
results for a Hydrogen chain in an electric field and for a
four-electron harmonium in a magnetic field. 
%We also verify that the
%corrected potential yields the atomic step
%structure in a perfect agreement with the original Becke-Johnson 
%approximation.
\end{abstract}

%\pacs{}

\maketitle

%\input{qdxholecurvature}

%\section{Introduction}

Electronic structure calculations are routinely carried
out by using density-functional theory~\cite{dft1,dft2} (DFT) and
its variants. 
The accuracy of a DFT calculation depends on 
the approximation used for the exchange-correlation energy
functional. Substantial efforts have been made in deriving
accurate approximations over the past few decades~\cite{dft3}.

Within spin-DFT (SDFT) the optimized-effective-potential (OEP) 
method~\cite{oep1,oep2,oep_review} provides 
a rigorous access to the {\em exact exchange} (EXX) energy,
%\begin{widetext}
\begin{equation}
E_{x}[\rho_{\sigma}] = -\frac{1}{2} \sum_{\sigma=\uparrow,\downarrow} \sum_{j,k=1}^{N_\sigma} 
\int d^3 r \int d^3 r'
\frac{\varphi_{j \sigma}^*(\vr)\varphi_{k \sigma}^*(\vr')
\varphi_{j \sigma}(\vr')\varphi_{k \sigma}(\vr)}{|\vr
- \vr'|} \; ,
\label{Ex}
\end{equation}
%\end{widetext}
and to the Kohn-Sham (KS) exchange potential $v_{x\sigma}(\vr)=\delta E_x/\rho_\sigma(\vr)$. 
Hartree atomic units (a.u.) are used throughout.
Above $\varphi_{j \sigma}(\vr)$ are the spin-dependent Kohn-Sham (KS) orbitals, 
with energies $\varepsilon_{j \sigma}$, and 
\begin{equation}
\rho_{\sigma}(\vr) = \sum_{j=1}^{N_{\sigma}}\left| \varphi_{j \sigma}(\vr) \right|^2
\end{equation}
is the ground-state density.
The OEP method leads to an integral equation which can be solved
iteratively together with the standard KS equations.
The main origin of practical complexity in the OEP method are the orbital shifts
containing unoccupied KS orbitals~\cite{grabo}. Despite great progress
in solving the equations
for various systems~\cite{IvanovHirataBartlett:99,Goerling:99,StaedeleMajewskiVoglGoerling:97,hrp,rp,Sharma1,Sharma2,demo1,demo2},
and in algorithmic developments~\cite{KuemmelPerdew:03,KuemmelPerdew:03-2},
efficient and accurate approximations reducing the
numerical burden of the full OEP scheme are still needed.

A simple approximation for the exchange potential has been
proposed by Becke and Johnson (BJ)~\cite{BJ}:
\bea
v^{\rm BJ}_{x\sigma}(\vr) & = & v^{\rm SL}_{x\sigma}(\vr) + \Delta v^{\rm
  BJ}_{x\sigma}(\vr) \nonumber \\
& = & v^{\rm SL}_{x\sigma}(\vr) + 
C_{\Delta v} \left[ \frac{ \tau_{\sigma}(\vr) }{ \rho_\sigma(\vr) }  \right]^{1/2},
\label{BJ_formula}
\eea
where
\be
v^{\rm SL}_{x\sigma}(\vr) = -\frac{1}{\rho_\sigma(\vr)} \sum_{j,k=1}^{N_\sigma} 
\int d^3 r' \frac{\varphi_{j \sigma}^*(\vr)\varphi_{k \sigma}^*(\vr')
\varphi_{j \sigma}(\vr')\varphi_{k \sigma}(\vr)}{|\vr
- \vr'|} \; ,
\label{slater}
\ee
is the Slater potential,  
\be\label{stau}
\tau_\sigma(\vr)=\sum_{j=1}^{N_\sigma} |\nabla\varphi_{j\sigma}(\vr)|^2
\ee
is (twice) the spin-dependent kinetic-energy density, and $C_{\Delta v} = \left[ 5/(12\pi^2)\right]^{1/2}$.
Note that the Slater potential in Eq.~(\ref{slater}), which is now 
the only numerical bottleneck, could be alternatively
approximated by the semi-local Becke-Roussel approach~\cite{becke_roussel}.
Interestingly, the {\em exact} exchange potential could be written
as in Eq.~(\ref{BJ_formula}), but with 
$\Delta v_{x \sigma}^{\rm BJ}(\vr)$ replaced by
$\Delta v_{x \sigma}^{\rm OEP}(\vr)$, which can be decomposed into 
the so-called Krieger-Li-Iafrate (KLI) approximation~\cite{KLI}, 
plus another correction given in terms of the 
orbital shifts~\cite{grabo}. As it is well known, in many cases the KLI approximation
is in good agreement with the full OEP.

Despite the semi-locality of $\Delta v^{\rm  BJ}_{x\sigma}(\vr)$
in Eq.~(\ref{BJ_formula}), the BJ potential is able to correctly
yield the step structure in the exchange potential of several 
atoms~\cite{BJ}. Moreover, it has recently been shown that the
BJ potential correctly reproduces the derivative discontinuity
for fractional particle numbers~\cite{armiento}.
During the first few years after its introduction, the BJ 
approximation has already been applied to various
systems~\cite{armiento,Staroverov,Kodera,Gaiduk,Fabien,Naoto,Tran}.
Impressively, the band gap of a large variety of extended systems
is extremely well reproduced~\cite{Fabien,Tran}.
However, as we will demonstrate below, the BJ potential may dramatically
fail in the presence of an electric or magnetic field, or a 
non-Coulombic external potential.

The limitation of the BJ potential originates from two facts: (i) it is not 
gauge-invariant and (ii) it is not exact for all possible one-electron systems.
These two problems may be fixed in similar fashion as demonstrated in our
recent derivation of a BJ-type approximation for two-dimensional 
systems~\cite{BJ_2D}. 

Before proceeding further, we would like to comment on the 
gauge-invariance requirement. For systems acted upon an external 
vector potential, the exchange potential we propose in this work should be 
identified as an approximation derived for 
the exact exchange-potential obtained within current-spin-density 
functional theory~\cite{vignale1,vignale2} (CSDFT) by taking the 
functional derivative of the exchange-correlation energy 
functional (written in terms of the spin-particle and vorticity
density) at constant vorticity.
On the other hand, it is clear that since $E_{x}$ in SDFT 
depends only on $\rho_{\sigma} (\vr)$, it must be a gauge-invariant 
quantity by definition. As a direct consequence, the 
corresponding $v_{x \sigma}(\vr)$ is gauge-invariant as well. 

Therefore, we propose the following correction
\bea\label{key}
v_{x \sigma}(\vr) & = & v^{\text{SL}}_{x \sigma}(\vr) + \Delta{v}^{\text{C}}_{x \sigma}(\vr) \nonumber \\
& = & v^{\text{SL}}_{x \sigma}(\vr) + C_{\Delta v}
\left[ \frac{ D_{\sigma}(\vr) }{ \rho_\sigma(\vr)}  \right]^{1/2}  \; ,
\eea
with
\be\label{D}
D_{\sigma}(\vr)= \tau_{\sigma}(\vr)-\frac{1}{4}\frac{\left( \nabla \rho_\sigma(\vr)
\right)^2}{\rho_\sigma(\vr)}-\frac{\vj^2_{p \sigma}(\vr)}{\rho_\sigma (\vr)} \; ,
\ee
where
\begin{equation}\label{current}
\vj_{p \sigma}(\vr)=\frac{1}{2i}\sum_{j=1}^{N_\sigma} \left[
 \varphi^*_{j \sigma}(\vr) \left(\nabla \varphi_{j \sigma}(\vr)\right) - \left(\nabla \varphi^*_{j \sigma}(\vr)\right) 
\varphi_{j \sigma}(\vr) \right]
\end{equation}
is the spin-dependent paramagnetic current density. The above
potential in Eq.~(\ref{key})  has a set of desirable properties listed below.

\begin{itemize}

\item In contrast with $\tau_{\sigma}(\vr)$, as it appears in the BJ expression,
the combination $\tau_{\sigma}(\vr) - \vj_{p
  \sigma}^2(\vr)/\rho_{\sigma}(\vr)$ is clearly gauge-invariant.~\cite{tao1,tao2,gamma}
As a result, also the corresponding potential is gauge-invariant.

\item In contrast with $\Delta v^{\rm  BJ}_{x\sigma}$, 
$\Delta{v}^{\text{C}}_{x\sigma}$ is zero for {\em all} one-particle systems.
This is easy to see by considering an arbitrary one-particle system
with $\rho_{\sigma}(\vr)= |\varphi_{\sigma}(\vr)|^2$ and
$\varphi_{\sigma}(\vr)=\sqrt{\rho_{\sigma}(\vr)}e^{i\theta(\vr)}$,
so that $D_{\sigma} (\vr) \equiv 0$ follows immediately from Eq.~(\ref{D}).
Alternatively, this may been seen by using the definition of $\tau_{\sigma}(\vr)$
and $\vj_{p \sigma}(\vr)$ [Eqs.~(\ref{stau}) and (\ref{current}), respectively]
in terms of $\varphi_{\sigma}(\vr)$, 
and by re-expressing $\rho_{\sigma}(\vr)= |\varphi_{\sigma}(\vr)|^2$
in the second term of $\Delta{v}^{\text{C}}_{x\sigma}$.

\item The asymptotic limit is correct for {\em any} $N$-electron finite system:
$\Delta{v}^{\text{C}}_{x\sigma}(\vr \rightarrow \infty)\rightarrow 0$
and then $v_{x
\sigma}(\vr \rightarrow \infty) \rightarrow v_{x \sigma}^{\text{SL}}(\vr
\rightarrow \infty) \rightarrow -1/r$.
In that limit all the terms in $D_\sigma$ are dominated by the highest 
occupied orbital~\cite{degeneracy},
and thus the system effectively behaves like a one-particle system
(see the preceding point). Below we discuss the asymptotic limit
in detail for two particular systems.

\item Equation~(\ref{key}) is consistent with the
limit of the homogeneous 3D electron gas (3DEG):
$\Delta v^{\text{OEP}}_{x \sigma} = 
\Delta v^{\text{BJ}}_{x \sigma} = 
\Delta{v}^{\text{C}}_{x \sigma} =
\left[ 3 \rho_{\sigma}/(4\pi) \right]^{1/3}$.

\item Calculation of $\Delta{v}^{\text{C}}_{x \sigma}$
instead of $\Delta v^{\text{BJ}}_{x \sigma}$ does not bring
any extra computational burden. 

\item Also, it is reassuring to note that, 
the exchange potential in Eq.~(\ref{key})
scales linearly as the exact one (see Appendix)~\cite{levyperdew}.

\end{itemize}

Finally we point out that the key object in the corrected
exchange potential, $D_{\sigma}(\vr)$, is familiar from various 
concepts in the literature. First, it is an important part of the
electron localization function~\cite{elf1,elf2,elf3}, and second, 
it enters in the expression of the local curvature of the exchange 
hole~\cite{dobson}. In the latter case, it is a part of the 
current-generalized forms~\cite{becke_canada,x1,c1} of the 
Becke-Roussel and Becke models for the
exchange~\cite{becke_roussel} and correlation~\cite{becke_correlation},
respectively.

Next we test our exchange potential against the KLI~\cite{KLI}, BJ, and
local-density approximation (LDA) for two different 
systems. We perform the self-consistent
KLI calculations applying the {\tt octopus}~\cite{octopus} 
DFT code on a real-space grid. The resulting KS orbitals
are then used as inputs in the approximations for the exchange
potentials. 

\begin{figure}
\includegraphics[width=0.80\columnwidth]{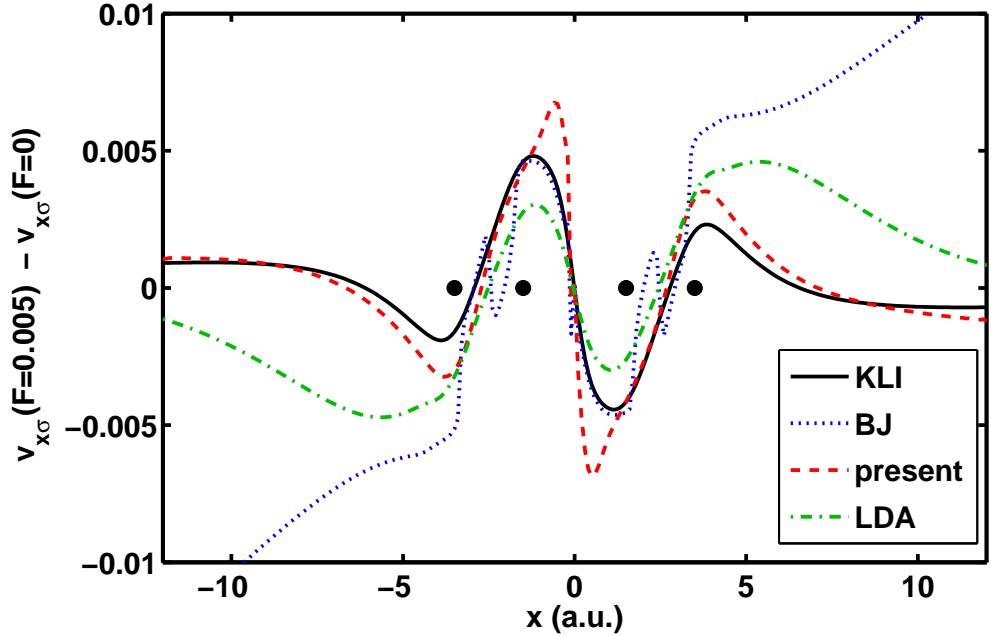}
\caption{(Color online) Difference in the exchange potentials 
for a four atom hydrogen chain with and without the linear electric field.
Black dots denote the positions of Hydrogen atoms. 
}
\label{fig1}
\end{figure}

First we consider a $H_4$ chain in an 
external linear field with the same system parameters as in
the work by Armiento, K\"ummel, and K\"orzd\"orfer~\cite{armiento} (AKK).
The system consists of two
$H_2$ ``molecules'' with an interpair distance of 2 a.u. 
separated by 3 a.u. The strength of the electric field, applied along
the $x$ direction,
is $F=0.005$ a.u. (hartree/bohr). Figure~\ref{fig1}
shows the {\em difference} in the exchange potential with
and without the electric field, i.e., $v_{x\sigma}(F=0.005)-v_{x\sigma}(F=0)$.
Overall, we find excellent agreement between KLI (solid line)
and the present approximation (dashed line) 
in Eq.~(\ref{key}). 

Figure~\ref{fig1} can be directly compared
to Figs. 3 and 5 in Ref.~\cite{armiento}.
We find the same
divergence in the BJ potential in the asymptotic
regime, as well as the large deviation of the LDA from the
KLI result. More importantly, we find that our corrected formula is 
closer to KLI as the potential proposed by AKK, i.e.,
\bea
&&v_{x \sigma}^{\text{AKK}}(\vr)  =  v^{\text{SL}}_{x \sigma}(\vr) + \nonumber \\
&& C_{\Delta v} \left(
\sqrt{ \frac{ \tau_{\sigma}(\vr) }{ \rho_\sigma(\vr)}}-\sqrt{-2 \varepsilon_{N_\sigma\sigma}} - \frac{x \; F}
{\sqrt{-2 \varepsilon_{N_\sigma\sigma}}} 
\right) \; ,
\label{AKK}
\eea
with $\varepsilon_{N_\sigma\sigma}$ corresponding to the energy of the highest occupied
KS orbital~\cite{factor}.
In particular, the AKK potential difference is close to zero on
the left ($x\lesssim -10$), whereas the KLI and the present 
approximation yield a finite value in that regime. 
Close to the atoms we find some
overestimation in the maxima (and minima) of our potential,
but, on the other hand, our approximation is free from
sharp kinks present in the BJ and AKK potential differences.

Next we compare in detail the asymptotic limit of the above example
given by the AKK potential [Eq.~(\ref{AKK})] and our potential [Eq.~(\ref{key})], respectively.
For atomic systems in the absence of external fields the 
asymptotic behavior of the KS orbitals is given by~\cite{grabo}
\be \label{aKS}
\varphi_{j \sigma}(\vr) \xrightarrow{\vr \rightarrow \infty} \Phi_{j \sigma}(r)f_{j \sigma}(\Omega) \; ,
\ee
with $\Phi_{j \sigma}(r)$ being the asymptotic radial 
wavefunction, and $f_{j \sigma}(\Omega)$ its corresponding angular
component. The asymptotic form of $\Phi_{j \sigma}(r)$ is~\cite{grabo}
\be \label{arKS}
\Phi_{j \sigma}(r) \xrightarrow{r \rightarrow \infty} r^{1/\beta_{j \sigma}} \frac{e^{-\beta_{j \sigma}r}}{r} \; ,
\ee
with $\beta_{j \sigma}=\sqrt{-2 \; \varepsilon_{j \sigma}}$. 
Substituting Eqs.~(\ref{aKS}) and (\ref{arKS}) into Eq.~(\ref{key}),
it is easy to find that the leading correction in the asymptotic
limit is given by
\bea \label{akey}
v_{x \sigma}(\vr) & \xrightarrow{\vr \rightarrow \infty} & v^{\text{SL}}_{x \sigma}(\vr) \nonumber \\
& + & C_{\Delta v}
\left[ \frac{ \tau_{\sigma}(\vr) }{ \rho_\sigma(\vr)} - (-2 \, \varepsilon_{N_\sigma\sigma})  + g_{N_\sigma\sigma}(\Omega)  \right]^{1/2}\, ,
\eea
where $g_{N_\sigma\sigma}(\Omega)=-\,[\nabla f_{N_\sigma\sigma}(\Omega)][\nabla f^*_{N_\sigma\sigma}(\Omega)]/|f_{N_\sigma\sigma}(\Omega)|^2$ is a purely angular term with
contributions coming from the second and third terms in $D_\sigma$ 
defined in Eq.~(\ref{D}). Making the subsequent asymptotic
expansion of $\tau_\sigma(\vr)/\rho_\sigma(\vr)\xrightarrow{\vr \rightarrow \infty} -2\epsilon_{N_\sigma\sigma}
-\,g_{N_\sigma\sigma}(\Omega)$, it is apparent that the term inside the square-root
in Eq.~(\ref{akey}) vanishes identically in the asymptotic regime.
We emphasize that, Eq.~(\ref{akey}) is similar, but not identical, to the AKK potential in
Eq.~(\ref{AKK}), with $F=0$. The main difference is that while the satisfaction
of several exact constraints (as explained above) enforces us to have all
the contributions of $D_\sigma$ {\em inside} the square-root, the correction in the AKK potential 
that enforces the vanishing of the exchange potential in
the asymptotic limit ($\sqrt{-2\epsilon_{N_\sigma\sigma}}$) is {\em outside}
the square-root.

Similar considerations apply when the external electric field $F$ is present.
Following AKK, in this case the asymptotic behavior of the KS
atomic orbitals along the direction of the applied electric field is given by~\cite{Airy}
\be
\varphi_{j \sigma}(x) \xrightarrow{x \rightarrow \infty} \eta^{-1/4}e^{-2\eta^{3/2}/3} \; ,
\ee 
with $\eta = (2F)^{1/3}(x-\varepsilon_{j \sigma}/F)$. Proceeding with the evaluation of 
$\Delta v_{x \sigma}^{\text{C}}(x \rightarrow \infty)$, we find
\be
\Delta v_{x \sigma}^\text{C}(x) \xrightarrow{x \rightarrow \infty} C_{\Delta v}
\left[ \frac{\tau_{\sigma}(x)}{\rho_{\sigma}(x)}-2(xF-\varepsilon_{N_\sigma\sigma}) \right]^{1/2} \; ,
\ee
which again is similar, but not identical, to Eq.~(\ref{AKK}). 
Making them identical would require an expansion of the 
argument inside the square-root, but this is unjustified, 
since in the asymptotic regime both contributions
are equally important. That is, 
$\tau_{\sigma}(x)/\rho_{\sigma}(x) \xrightarrow{x \rightarrow \infty} 2(xF-\varepsilon_{N_\sigma\sigma})$,
leading to a cancellation of both terms inside the square-root.

As a conclusion of the analysis in two previous paragraphs,
let us emphasize that {\it both} our $v_{x \sigma}(\vr)$ and 
$v_{x \sigma}^{\text{AKK}}(\vr)$ reproduce the correct asymptotic
limit of the exact exchange potential, but in different ways.
On the other side, $v_{x \sigma}^{\text{AKK}}(\vr)$ is
system-dependent, being only 
valid for atomic systems in presence of a bias,
while our exchange potential is system-independent, 
being valid for {\it any} 3D system,  
in the presence of any electric and/or magnetic fields
(see below). Also it is better suited for dealing with currents. 
Moreover, our potential does not require explicit knowledge of
external fields (as it should be for any
standard density functional) and/or KS eigenenergies -- only (occupied) 
KS orbitals are needed.

\begin{figure}
\includegraphics[width=0.70\columnwidth]{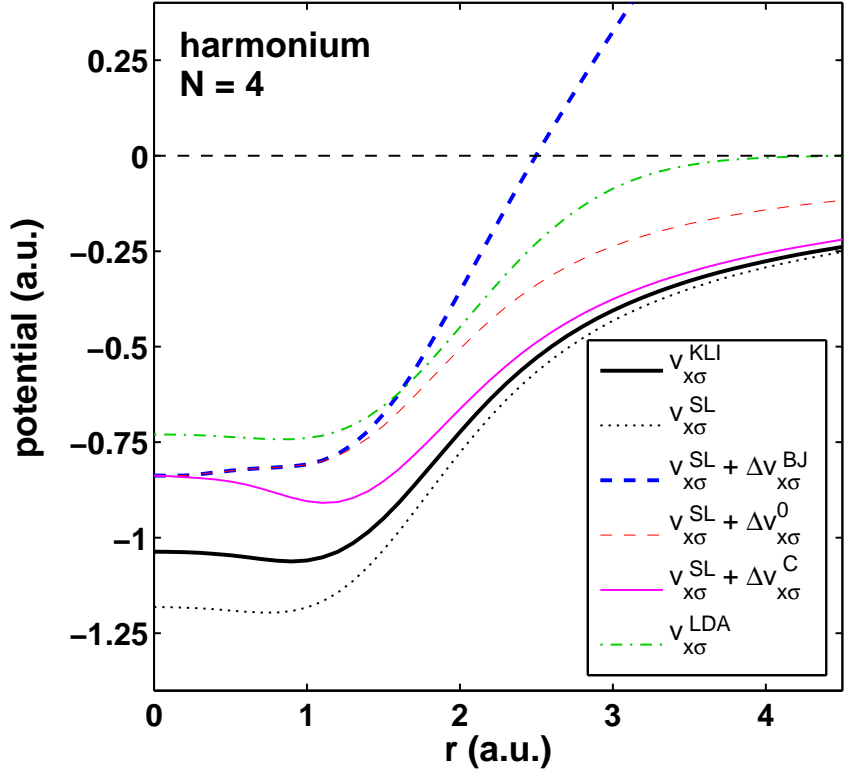}
\caption{(Color online) Exchange potentials for a four-electron
harmonium in external magnetic field $B=300$ a.u.
}
\label{fig2}
\end{figure}

In Fig.~\ref{fig2}
we show the exchange potentials for a fully spin-polarized 
four-electron ``harmonium'', i.e., a 3D harmonic oscillator 
with a radial external potential
$v_{\rm ext}(r)=\omega^2 r^2/2$, where $\omega=1$ a.u. 
This type of potential
could be used as a realistic model for
quantum dots, i.e., electrons confined in
atomic clusters or semiconductor heterostructures.  
We have also set an external, uniform magnetic field to
$B= 300$ a.u., 
so that the occupied KS states have angular momenta
$l=0,-1,-2,-3$ and hence there are orbital currents in the system.
The BJ potential (thick dashed line) shows erroneous divergent 
behavior. Similar divergence appearing in two-dimensional harmonic
oscillator has been analyzed in detail in Ref.~\cite{BJ_2D}.
It was shown that the linear increase in the BJ potential
at large $r$ follows directly from the asymptotic limit 
of the single-particle wave functions, which, in the case of a 
parabolic confining potential, decays
as $\exp(-r^2)$ in contrast with the atomic wave
function that decays as $\exp(-r)$. The situation
is the same in the 3D case considered here.
The LDA result (dash-dotted line in Fig.~\ref{fig2}), 
on the other hand, largely
underestimates the exchange potential throughout the system.
Similar tendency is shown by the BJ potential modified by the
gradient term [second term in Eq.~(\ref{D})] but {\em without} the
current term  [third term in Eq.~(\ref{D})], i.e., without
enforcing the gauge-invariance (thin dashed line marked
by $v^{\rm SL}_{x\sigma}+\Delta v_{x\sigma}^0$).
The closest resemblance of the KLI potential in Fig.~\ref{fig2}
is clearly given by the present approximation in Eq.~(\ref{key}). 
Also it can observed that although the asymptotic limit
is very well reproduced, close to the core of 
this system we still find some deviation, both for the BJ and 
for our corrected exchange potential. This gives evidence that
further improvements may be suggested in future works.

\begin{figure}
\includegraphics[width=0.70\columnwidth]{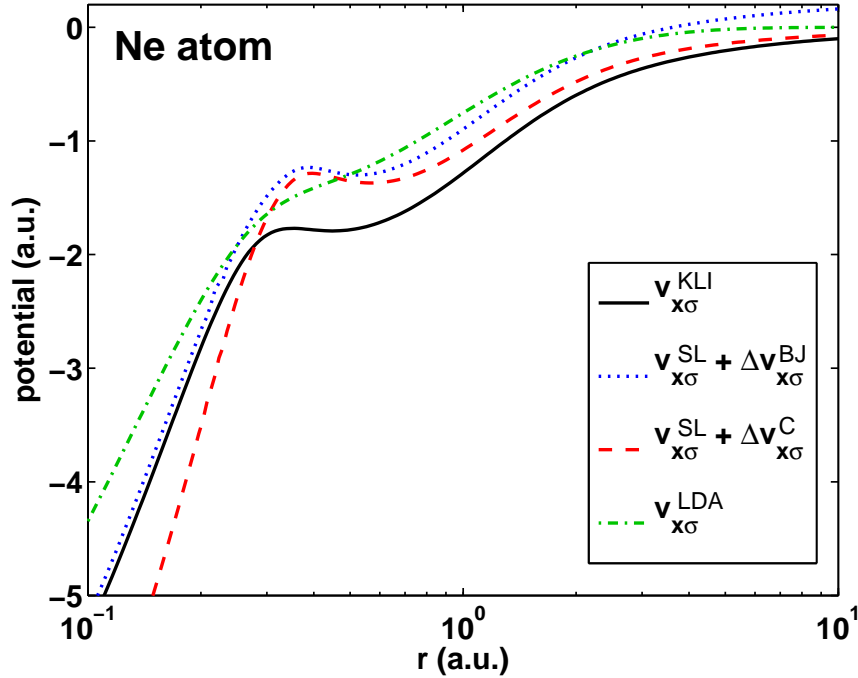}
\caption{(Color online) Exchange potential for a closed-shell neon
  atom in the ground state.
}
\label{fig3}
\end{figure}

Finally, we verify that the atomic step structure at electronic 
shells -- one of the motivations behind the original Becke-Johnson 
approximation~\cite{BJ} -- is reproduced by the corrected potential.
In Fig.~\ref{fig3}
we show the exchange potential for a closed-shell neon atom in
its ground state with no external fields present. The corrected
potential (dashed line) has the step structure at $r\sim 0.3$ a.u. in 
agreement with the BJ potential (dotted line) and with the KLI
potential (solid line). The OEP solution shown in Ref.~\cite{BJ}
shows a slightly sharper shoulder than the KLI one, but generally they
are very similar, which is in accordance with previous studies in the 
literature (see, e.g., Ref.~\cite{grabo}). In the exchange-LDA 
potential (dash-dotted line) the step (or shoulder) structure is 
missing. 
Note that for consistency with the previous results we have not 
imposed a shift to the BJ potential, which is 
a prerequisite having different definitions~\cite{BJ,armiento}.
Secondly, we point out that Fig.~\ref{fig3} results from 
self-consistent calculations for all potentials, respectively.
The difference from the non-self-consistent results, obtained by
using the KLI orbitals as the input, was found to be negligible.

We point out that for atoms at small $r$  
our potential decreases more strongly than the KLI (or BJ)
potential (see Fig.~\ref{fig3}). 
In fact, at $r=0$ the correction term $\Delta v_x^{\rm C}$
is significantly smaller than $\Delta v_x^{\rm KLI}$. 
This difference is not present at $r=0$
in a harmonic confinement discussed above, since in that
case it is easy to show that the correction terms in the BJ and
our potential have exactly the same value due to the Gaussian
form of the orbitals;  this feature can also be observed directly 
from Fig.~\ref{fig2}.
Nevertheless, the deviation found
in the atomic case close to the nuclei suggests that an additional 
effort beyond the present contribution may be required.

Concluding, we have seen that Becke-Johnson potential
may dramatically fail when considering systems in external electric
and magnetic fields. We have proposed a universal 
correction which is gauge-invariant for complex Kohn-Sham 
orbitals and exact for any one-particle system.
The improved approximation is suited for dealing with 
with current-carrying states, and it also
recovers the correct asymptotic behavior of the exact exchange
potential for any many-electron system.
We have demonstrated the very good performance of the resulting exchange
potential by considering a hydrogen chain in an external electric
field as well as a four-electron harmonic oscillator in a magnetic field. 

\begin{acknowledgments}
We would like to thank Giovanni Vignale for useful discussions.
This work was supported by the Academy of Finland, Deutsche 
Forschungsgemeinschaft, and the EU's Sixth Framework
Programme through the ETSF e-I3.
C.R.P. was supported by the European Community through a Marie 
Curie IIF (Grant No. MIF1-CT-2006-040222).
S.P. acknowledges support by DOE grant DE-FG02-05ER46203.
\end{acknowledgments}

\appendix* 

\section{}
We may write the term beyond the Slater contribution 
to the exchange potential as
\be\label{MBJ3}
\Delta {v}^{\text{C}}_{x \sigma}(\vr) =
 C_{\Delta v} \left[ \frac{ D_{\sigma}(\vr) }{ \rho_\sigma(\vr)}   \right]^\alpha \; ,
\ee
and determine $\alpha$ under the constraint of exact linear 
scaling~\cite{levyperdew}.
Under uniform scaling of the coordinates, $\vr\rightarrow\lambda\vr$, the norm-preserving many-body wavefunction is given by
$\Psi_\lambda(\vr_1,...\vr_N)=\lambda^{3N/2}\Psi(\lambda\vr_1,...,\lambda\vr_N)$ (with $0 < \lambda
< \infty$). As a consequence, the 3D density scales with $\lambda$ as follows:
$\rho_{\sigma}(\vr) \rightarrow \lambda^3 \rho_{\sigma}(\lambda \vr)$. This leads to the result that
the KS orbitals in 3D are seen to scale as
$\varphi_{j \sigma}(\vr) \rightarrow \lambda^{3/2} \, \varphi_{j \sigma}(\lambda \vr)$. Thus,
$\tau_{\sigma}(\vr) \rightarrow \lambda^5 \, \tau_{\sigma}(\lambda \vr)$,
$\nabla \rho_{\sigma}(\vr) \rightarrow \lambda^{4} \, \nabla_{\lambda \vr} \rho_{\sigma}(\lambda \vr)$,
and $\vj_{p \sigma}(\vr) \rightarrow \lambda^4 \, \vj_{p \sigma}(\lambda \vr)$.
Substituting these relations into Eq.~(\ref{MBJ3}) yields
$\Delta {v}^{\text{C} \, \lambda}_{x \sigma}(\vr) =
\lambda^{2 \alpha} \Delta {v}^{\text{C}}_{x \sigma}(\lambda \vr)$,
which fulfills the linear scaling constraint {\it only} if 
$\alpha = 1/2$ in agreement with the expression in Eq.~(\ref{key}).


\begin{thebibliography}{ll}
\bibitem{degeneracy} To simplify the analysis we assume that the 
highest occupied KS orbitals are non-degenerate.
\bibitem{factor}
Note that our definition of $\tau_{\sigma}(\vr)$ includes an extra factor of two.
\bibitem{Airy}
The decay of the KS orbitals in the direction of the external electric field is given by the asymptotic
decay of the Airy functions.
\end{thebibliography}
\end{document}